\documentclass[preprint,preprintnumbers,amsmath,amssymb,amssymb]{revtex4}
\usepackage{graphicx}
\usepackage{dcolumn}
\usepackage{bm}
\begin{document}

%
%



\newcommand{\ttbs}{\char'134}


\title{Possibilities for charged Higgs bosons at the LHC in a SU(3)$_L\otimes$U(1)$_N$ Model}
\author{J.\ E.\ Cieza Montalvo$^1$, Nelson V. Cortez Jr.$^2$,
J. S\'a Borges$^1$ and Mauro D. Tonasse$^3$}
\affiliation{$^1$ Instituto de F\'{\i}sica, Universidade do Estado do Rio
de Janeiro, Rua S\~ao Francisco Xavier 524, 20559-900 Rio de Janeiro,
RJ, Brazil\\
$^2$ Rua Justino Boschetti 40, 02205-050 S\~ao Paulo, SP, Brazil\\
$^3$ {\it Campus} Experimental de Registro, Universidade Estadual Paulista, Rua
Tamekishi Takano 5, 11900-000, Registro, SP, Brazil}  
\begin{abstract}
We have studied the branching ratios of doubly charged Higgs bosons at the LHC using a version of  the SU(3)$_L\otimes$U(1)$_N$ electroweak model. At the end of this work we have made a very simple plotting comparating the total cross section of this model using Drell-Yan, gluon-gluon fusion and Left-right symmetric model. 
\end{abstract}
\maketitle
Doubly charged scalar particles arise in many scenarios extending the weak interactions beyond the Standard Model (SM) as the left-right symmetric model (LR) \cite{pati} and Higgs triplet models \cite{geor}. In the LR Model electroweak theory such a particle is a member of a triplet Higgs representation which plays a crucial part in the model. The gauge symmetry of the LR Model  is broken to the SM symmetry due to a non-vanishing expectation value in the vacuum of the neutral component of the triplet right-handed Higgs. The Left-Right Model predicts two kinds of doubly charged particles with different interactions. \par
\begin{table}[h]
\caption{The approximate values of the masses (in GeV) used in the present work.}
\label{table:1}
\begin{tabular}{c|cccccccccccc}
\hline
$f$&$m_E$&$m_M$&$m_T$&$m_{H_1^0}$&$m_{H_2^0}$ & $m_{H_3^0}$ & $m_{h^0}$ & $m_{H^\pm_1}$ & $m_{H^\pm_2}$ & $m_V$ & $m_U$ & $m_{Z^\prime}$\\
\hline
$\approx 0$ & 194 & 1138 & 2600 & 874  & 1322  & 2600 & 0 & 426 & 1315 & 603 & 601 & 2220\\
-99.63 & 194 & 1138 & 2600 & 874 & 1322 & 2600 & 520 & 218 & 1295 & 603 & 601 & 2220\\
\hline
\end{tabular}
\end{table}
The so called 3-3-1 Models \cite{pisa,vice} are another interesting class of electroweak models that also predict such particles. It is able to solve the fermion family's replication problem through a simple relation between the colors number and the anomaly cancellation mechanism. In a similar fashion as occurs in LR Model, the seesaw mechanism can be incorporated in some versions of the 3-3-1 Models \cite{nelson}. \par 
Since the 3-3-1 Models are good candidates to physics beyond the SM, it is interesting to evaluate if the future accelerators will produce events in sufficient numbers to detected
some of the 3-3-1 Higgs bosons. In particular, there is an increasing interest in the phenomenology associated with doubly charged Higgs bosons \cite{some}. Here we are interested in one of such version of the 3-3-1 Models for which the scalar fields come only in triplet representation \cite{vice,nelson} of the  $SU(3)$ gauge group. \par
It predicts four neutral $\left(H_1^0, H_2^0, H_3^0, h0\right)$, four single charged $\left(H_1^\pm, H_2^\pm\right)$ and two doubly charged $\left(H^{\pm\pm}\right)$ Higgs bosons. In the gauge sector, beyond the standard gauge fields, the model predict also the extra neutral $Z^\prime$ and the charged  $V^\pm$ and $U^{\pm\pm}$ bosons. The fermionic sector is enlarged by the heavy leptons $E^\pm$, $M^\pm$ and $T^\pm$ and the quarks $J_1$, $J_2$ and $J_3$. \par
\begin{table}[h]
\caption{Branching ratio for the $H_3^0$ decays with $m_{H^0_3} = 2600$ GeV. Here $BR^0_{XY}$ stands for $BR\left(H_3^0 \to XY\right)$.}
\label{table:2}
\begin{center}
\begin{tabular}{c|cccccc}
\hline
$f$ (GeV) & $BR^0_{H_1^0H_20}$ & $BR^0_{H_1^+H_1^-}$ & $BR^0_{H_2^+H_2^-}$ & $BR^0_{H_1^+H_2^-}$ & $BR^0_{E^+E^-}$ & $BR^0_{M^+M^-}$ \\
\hline
$\approx 0$ & 3.35$\times 10^{-5}$ & 2$\times 10^{-8}$ & No & 0.9999 & 2 $\times 10^{-7}$ & 2 $\times 10^{-6}$ \\ 
-99.63 & 3.14 & 4 & 4 $\times 10^{-7}$ & 0.9999 & 2 $\times 10^{-7}$ & 2 $\times 10^{-6}$ \\
\hline
\end{tabular}
\end{center}
\end{table}
\begin{figure}[t]
\includegraphics[width=35pc]{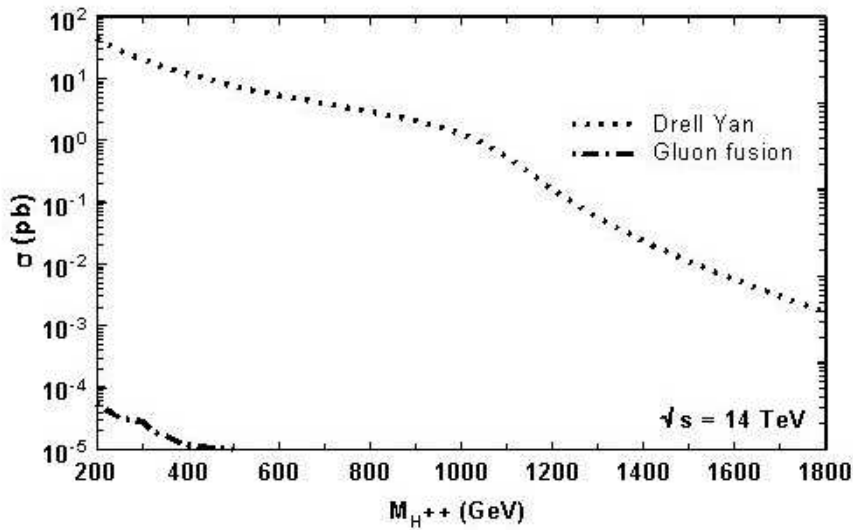}
\caption{Total cross section for the process $pp \to H^{--}H^{++}$ as a function of $m_{H^{\pm\pm}}$ for $f = 0$ GeV at $\sqrt{s} = 14$ TeV for Drell-Yan (dotted line) and Gluon-Gluon fusion (dot-dashed line).}
\end{figure}
The main mechanism for the production of Higgs particles in $pp$ collisions occurs in association with the bosons $\gamma$, $Z$, $Z^{'}$, $H_{1}^{0}$ and $H_{2}^{0}$ through the mechanism of Drell-Yan and with the $H_{1}^{0}$, $H_{2}^{0}$ and $H_{3}^{0}$ by gluon-gluon fusion. In all calculations in this work we are considering that the charged fermionic mixing matrices are diagonals. We first evaluate the differential cross section for the Drell-Yan, that is, the  process $pp \to H^{++}H^{--}$ which takes place through the exchange of refereed
bosons in the $s$-channel. \par
\begin{table}[h]
\caption{Branching ratios for $H^\pm\pm$ decays with $m_{H^{\pm\pm}} = 1309$ GeV. $BR^{\pm\pm}_{XY}$ stands for $BR^{\pm\pm}_{XY}= 10^3 \times BR\left(H^{\pm\pm} \to XY\right)$.}
\label{table:3}
\begin{center}
\begin{tabular}{c|ccccccc}
\hline
$f$ (GeV) & $BR^{\pm\pm}_{\bar J_1q}$ & $BR^{--}_{\ell^- E^-}$ & $BR^{--}_{\ell^- M^-}$ & $BR^{++}_{\ell^+ E^+}$ & $BR^{++}_{\ell^+ M^+}$ & $BR^{\pm\pm}_{U^{\pm\pm}\gamma}$  \\
\hline
$\approx 0$ & 0.001 & 0.08 & 0.005 & 3 & 6 & 29 \\
    -99.63  & 2 & 0.001 & 0.004 & 0.4 & 4 & 2 \\
\hline\hline
$f$ (GeV) & $BR^{\pm\pm}_{W^\pm H_2^\pm}$ & $BR^{\pm\pm}_{V^\pm H_1^\pm}$ & $BR^{\pm\pm}_{H_1^\pm H_2^\pm}$ & $BR^{\pm\pm}_{U^{\pm\pm}H_1^0}$ & $BR^{\pm\pm}_{U^{\pm\pm}Z}$ & $BR^{\pm\pm}_{U^{\pm\pm}h^0}$\\
\hline
$\approx 0$  & No & 19 & No & No & 444 & 2\\
    -99.63 & 0.09 & 13 & 329 & 6 & 146 & 0.5\\
\hline
\end{tabular}
\end{center}
\end{table}
\begin{figure}[t]
\vglue -0.85 cm
\includegraphics[width=35pc]{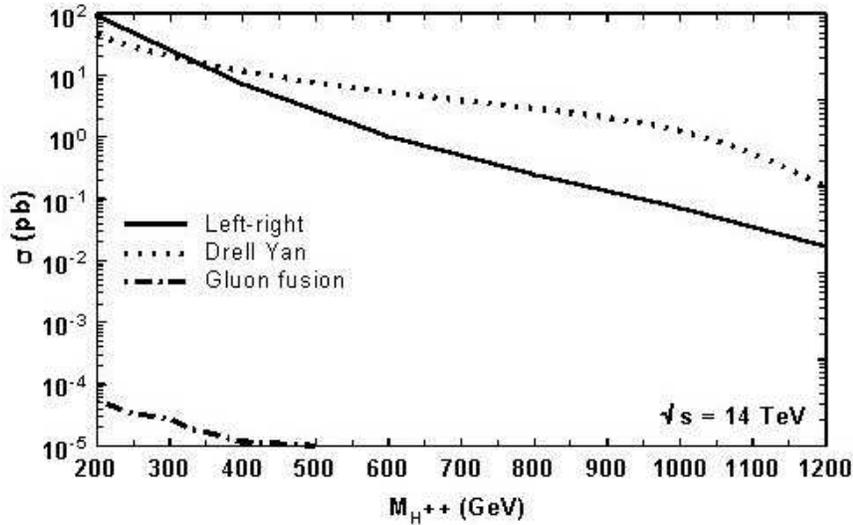}
\vglue -1 cm
\caption{Total cross section for the process $pp \to H^{--}H^{++}$ as a function of $m_{H^{\pm\pm}}$ for $f = 0$ GeV at $\sqrt{s} = 14$ TeV for Drell-Yan (dotted line) and Gluon-Gluon fusion (dot-dashed line) and Left-Right symmetric model (solid line).}
\end{figure}
We have considered two possibilities: $f \simeq 0$ and $f = -99.63$ GeV, where $f$ is the strenght of the trilinear coupling of the Higgs potential. The masses of the exotic bosons Table \ref{table:1} are in accordance with the estimates of the CDF and D0 experiments, which probes their masses in the 500 GeV $-$ 800 GeV range.\par
We calculate the doubly charged Higgs pair
production by computing the contributions due the Drell-Yan and
quark loop processes. We present, in Fig. 1, 
the cross section for the process $pp \to H^{--}H^{++}$ at the LHC energies (14 TeV) computed from the  Drell-Yan mechanism and from the gluon-gluon. One can observe that the cross section corresponding to the Drell-Yan mechanism is four orders of magnitude larger than the quark loop contribution. \par
Next we have computed the branching ratios for the $H_3^0$ decay with $m_{H_3^0}=2600$
GeV (Table \ref{table:2}). The branching ratios for $H^{\pm\pm}$ decay with $m_{H^{\pm\pm}} = 1309$ GeV are presented in Table \ref{table:3}. Finally the Fig. 2 show the comparison of our results with those obtainned from the Left-Right symmetric model \cite{gu}. 
As a result we show that the Drell-Yan mechanism contribute much more than the gluon-gluon fusion to the total cross section production. \par 
The analysis of these yields show that, although a large number of doubly charged Higgs can be produced by the Drell Yang mechanism, the decays of these particles into ordinary fermions do not lead to a good signature for its detection even for LHC energies. We can observe that the window for the free parameters is small, because of the constraints imposed on the model. 

\end{document}